\documentclass{ws-rv9x6}
\usepackage{amsmath}
\usepackage{subfigure}     
\usepackage{ws-rv-van}     
\makeindex
\usepackage{color}
\definecolor{red}{rgb}{1,0,0}
\def\mydoubleq#1{``#1''}

\begin{document}

\chapter[Transport and the first passage time problem with application to cold atoms in optical traps]{
Transport and the first passage time problem \\
with application to cold atoms in optical traps}
\label{ra_ch1}

\author[Eli Barkai and David A. Kessler]{Eli Barkai, David A. Kessler}

\address{Department of Physics\\
 Institute of Nanotechnology and Advanced Materials\\
 Bar-Ilan University, Ramat-Gan
52900, Israel}

\begin{abstract}
Measurements of spatial diffusion of cold atoms in optical lattices
have revealed anomalous super-diffusion, which is controlled by the depth of the
optical lattice. We use first passage time statistics to derive the diffusion
front of the atoms. In particular, the distributions
of areas swept under the first passage curve till its first arrival,
and of areas under the Bessel excursion are shown to be powerful
 tools in the analysis
of the atomic cloud.  A rather general relation between first passage time
statistics and diffusivity is discussed, showing that first passage
time analysis is  a useful tool in  the calculation of transport coefficients. 
A brief introduction to  the semi-classical description
of Sisyphus cooling is provided which yields a rich phase diagram 
for the dynamics.
\end{abstract}

\body

\section{Introduction}\label{ra_sec1}

 Fick's second law predicts how diffusion causes concentration
to change in time. Instead of concentration we will use a probability
density in one dimension $P(x,t)$ and then Fick's law reads
\begin{equation}
\frac{\partial P}{\partial t} = K_2 \frac{\partial^2 P}{ \partial x^2 } 
\end{equation} 
where $K_2$ is the diffusion coefficient. For an open system, and
particles initially localized at the origin, the solution is a spreading Gaussian
packet, with a width proportional to time: 
\begin{equation}
\langle x^2 \rangle = 2 K_2 t . 
\end{equation} 
Such normal diffusion is ubiquitous,
which is hardly surprising since the diffusion is merely a
sum of  random displacements with zero mean.
 Hence, following the central limit
theorem we expect to find  in experiment a Gaussian  distribution for the particle position.
The Gaussian universality of diffusion processes
 leaves us with one difficult task,
and that is to compute the diffusion constant $K_2$. This constant
is non-universal, unlike the nearly universal shape of the Gaussian
probability packet. Still, non-equilibrium statistical mechanics gives us
a recipe (in principle) for the calculation of $K_2$ which is the
famous Green-Kubo formula
\begin{equation}
K_2 = \int_0 ^\infty {\rm d} \tau \langle p(t+\tau) p(t) \rangle/ m^2 
\label{eqGK}
\end{equation} 
where $m$ is the mass of the particle. 
This relation between the momentum correlation function and the diffusivity
is extensively used in the context of hard and soft condensed matter.
The correlation function and the underlying process itself is assumed to
be stationary so $\langle p(t+\tau) p(t)\rangle$ is independent of $t$. 
The Green-Kubo formula is important since once we obtain the diffusivity
of a system we also know its mobility. This is made possible via linear
response theory and the well-known Einstein relation which connects
the diffusivity with the mobility via the thermal scale $k_b T$
where $T$ is temperature. 

 All this was established a long time ago,
 so it is not surprising that physicists have
turned their focus to problems that violate this normal behavior.
One example is the diffusion and transport of atoms in optical molasses 
driven by 
counter propagating laser beams \cite{Laser}.
 This system has attracted considerable attention,
both for applied purposes like cooling and control of atoms and from
a more fundamental point of view, since it  exhibits unusual
friction and in some cases large deviations from ordinary statistical
mechanics. Indeed laser cooling is today  the tool of choice in many
laboratories which investigate low temperature physics. So basic understanding
of statistical properties of laser cooled atoms is a timely subject.

 Within the semi-classical picture of Sisyphus cooling, the atoms are subject to a velocity 
dependent friction force, $F(v)$, defined more precisely below.  At low velocities,  $F(v) \propto - v$, which is 
of course normal in the sense that it mimics Stoke's friction for a macroscopic
Brownian particle in a viscous medium at room temperatures. Such a
friction force is intuitive since it indicates that the faster the particle
 moves
the more energy it dissipates to the surrounding bath. The large
$v$ behavior of the atomic friction on the other hand is counterintuitive:
it decreases with velocity, $F(v) \sim - 1/v$, thus when $v\to \infty$
the system becomes frictionless. Such a system can be called asymptotically
dissipationless, and it exhibits physical behaviors very different
than standard systems such as non-dissipative ones
and systems with kinetic dissipation which
does not vanish at high speeds.    

\begin{figure}\begin{center}
\includegraphics[width=0.8\textwidth]{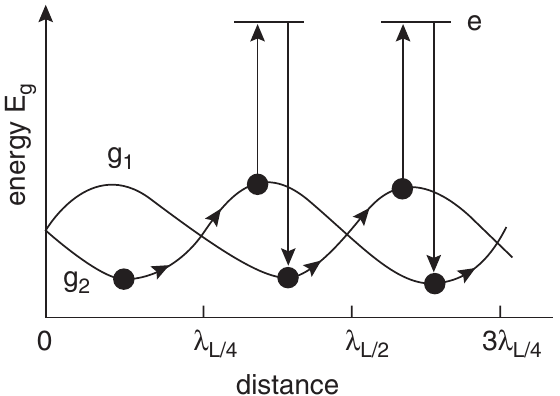}
\end{center}
\caption{
Sisyphus was punished by Greek gods to push a heavy stone
uphill just to have it roll back to its starting point. In Sisyphus
cooling, an atom will go up a potential hill created by laser fields, and
then falls down to the local ground state, thus loosing energy. Such a scheme
is made possible only when transitions from the potential maximum to minimum
 are statistically preferred, which in turn is made possible
by clever quantum mechanics, and polarization fields (which control the transitions) which are
periodic function of distance specially designed to correspond to the spatial
modulation of the energy surfaces (the period is determined  by the
wavelength of 
the laser field, $\lambda_L$).
 However, if Sisyphus rans like a bolt,  over the hills
and the valleys  he can fool the gods. Namely
Greek gods cannot push him down from the top of the hills, since they
need some finite time to implement his punishment. 
This corresponds to a fast atom whose transitions are not synchronized with
it being on maximum of energy surface. So for  fast particles, the Sisyphus
friction becomes small and decreases, according to the detailed calculations,
 like $F(v) \propto -1/v$.
For more  on Sisyphus cooling see \cite{Laser}.
}
\label{fig0}
\end{figure}

 This behavior of the friction force is due to the fact that the laser cooling is not effective
beyond some critical velocity. Briefly, the effect is caused since
a fast particle cannot distinguish the uphills from the downhills
in the spatially periodic optical lattice; hence the Sisyphus mechanism
which favors the loss of energy when the particle is close to the maximum 
of the periodic lattice is not effective (see Fig. \ref{fig0} and caption).
 Since the friction is
vanishingly small for large velocities the equilibrium distribution
of the velocities can be shown to obey power law statistics (see details below), and so various average quantities are
dominated by fast particles (compared for example with the Maxwellian distribution
which is Gaussian). More generally, the kinetics and equilibrium properties
of atoms under laser cooling are far from standard: anomalous diffusion,
non Gibbsian states, etc. can be found not only in Sysiphus cooling but also
in other cooling approaches like sub-recoil laser cooling \cite{Levybook}, and
in Doppler cooling as well. 

 Indeed in the closing paragraph of Cohen-Tannoudji's Nobel lecture~\cite{CTNobel}
he writes
\mydoubleq{It is clear finally that all the developments which have 
occurred in the field of laser cooling
 and trapping are strengthening the connections
which can be established between atomic physics and other branches of physics
such as condensed matter or statistical physics. The use of L\'evy statistics
for analyzing sub-recoil cooling is an example of such a fruitful dialogue}.
Here such a dialogue is extended to the popular Sysiphus cooling mechanism. 
L\'evy
statistics roughly refers to power law distributions, which violate
the conditions leading to the Gaussian central limit theorem. An example of this
is found in a recent experiment  performed at the Weizmann Institute.  

\subsection{Spatial diffusion of cold atoms}

 There, Sagi, et al.~\cite{Sagi} measured  the diffusionSisyphus
of ultra-cold $^{87}$Rb atoms in a one dimensional optical lattice.
Starting with a very narrow atomic cloud they recorded the time
evolution of the density
of the particles, here denoted $P(x,t)$ (normalized to unity).
Their work employed the well-known Sisyphus cooling scheme \cite{CT}.
As predicted theoretically  by Marksteiner, et al. \cite{Zoller},
the diffusion of the atoms was not Gaussian, so that the
assumption
that the diffusion process obeys the
standard  central limit theorem is not valid in this
case.  We recently determined  the precise
nature of the non-equilibrium spreading of the atoms, 
in particular the dynamical phase diagram
of the various different types of behaviors exhibited 
as the depth of the optical potential is varied, at least within the semiclassical approach.

In  Ref. \refcite{Sagi},
the anomalous diffusion data was compared to the 
solutions of the fractional diffusion equation \cite{Saichev,MBK,Review}
\begin{equation}
\frac{\partial^\beta  P(x,t) }{ \partial t^\beta} = K_\nu \nabla^\nu P(x,t),
\label{eq01}
\end{equation}
with $\beta = 1$,
so that the time derivative
on the left hand side is a first-order derivative. The fractional
space derivative on the right hand side is a Weyl-Rietz fractional
derivative~\cite{Review}, defined below. Here
the anomalous diffusion coefficient
 $K_\nu$ has units $\textrm{cm}^\nu/\textrm{sec}$. A fundamental challenge is to derive such fractional equations from
 a microscopic theory, without invoking power-law statistics in the first place.  Furthermore,
 the solutions of such equations exhibit a diverging mean-square displacement $\langle x^2 \rangle = \infty$,
 which violates the principle of causality\footnote{See the discussion in Ref. \refcite{KSZ} where the unphysical nature of L\'evy flights is discussed and the resolution in terms of L\'evy walks is addressed.}, which restricts physical phenomena to spread at finite
 speeds.  So how can fractional equations like Eq. (\ref{eq01}) describe physical reality?  We will address this paradox in this work.
The solution of Eq.
(\ref{eq01}) for an initial narrow cloud
is given in terms of a L\'evy distribution (see details below).
The L\'evy distribution generalizes the Gaussian distribution in the
mathematical problem of the sum of a large number of 
independent random variables symmetrically distributed about 0, in the case where the variance of
the summands diverges, corresponding physically to
scale free systems.
Here
our aim is to derive  L\'evy statistics and the fractional diffusion equation
from the semi-classical picture of Sisyphus cooling. Specifically, we will show that $\beta=1$
and relate the value of the exponent $\nu$ to the depth
of the optical lattice $U_0$, deriving an expression for the
constant $K_\nu$. Furthermore, we discuss the limitations of the
fractional framework, and show that for a critical value of
the depth of the optical lattice,
the dynamics switches to a non-L\'evy behavior (i.e. a regime where
Eq. (\ref{eq01}) is not valid); instead it is related to Richardson-Obukhov diffusion found in turbulence.
Thus the semiclassical picture predicts
a rich phase diagram for  the
atomistic
diffusion process. 
We will then compare the results of this analysis to the experimental
findings, and see that there are still unresolved discrepancies between the experiment and the theory.  Reconciling the two thus poses
a major challenge for the future.

 Since usual approaches to diffusion, namely Fick's second law and the
 Green-Kubo formula break down for the case of interest,
we cannot use ordinary approaches. Here the power of first passage time
statistics enters. We will show how analysis of first passage time
statistics yields the diffusion front of a packet of particles. In essence the
first passage time tool replaces the failing Green Kubo formula, and
leads us to the solution of the problem, namely  the exponents
$\beta$ and $\nu$ as well as the generalized diffusion constant $K_\nu$.
But this is a long journey, so first  let us briefly review the concept
of Brownian excursions, first passage times for simple Brownian
motion, and  the distribution of the area swept under Brownian
motion till its first passage. 

\subsection{Brief Survey: Joint PDF for first passage time and
the area swept under Brownian motion}

 Consider a Brownian motion in one dimension. The time $\tau_f$
 it takes
the particle starting at $x=\epsilon>0$ to reach $x=0$
for the first time is the first passage time.
The distribution  of $\tau_f$ has been
 well investigated both for Brownian motion and
for other types of random walks \cite{Schro,Redner}. 
A quantity which  until recently
was only of mathematical
 interest  is the 
total area  under the Brownian path in the time interval
$(0,\tau_f)$, which was treated by Kearney and Majumdar~\cite{Kearney}.
 This area,  which we denote $\chi_f$, is obviously positive if we start 
at $x=\epsilon>0$. Later we will take $\epsilon \to 0$ which is a subtle point,
but for now we keep $\epsilon$  finite. 
 The pair of random quantities $\chi_f$ and $\tau_f$ are clearly correlated
since a large $\tau_f$ implies statistically a large $\chi_f$.
The joint probability density function is denoted as $P(\tau_f,\chi_f)$.
Using Bayes' law we may write this density as
\begin{equation}
\psi(\tau_f,\chi_f) = g (\tau_f) P(\chi_f|\tau_f)  
\label{eqJoint}
\end{equation}
where $g(\tau_f)$ is the first passage time probability density function
(PDF), which at least for Brownian
motion is well investigated. In particular, $g(\tau_f)$
decays as a power for large $\tau_f$,
\begin{equation}
g(\tau_f) \propto (\tau_f)^{-3/2},
\label{eqfpS}
\end{equation}
so that  the average  first passage time,
 $\langle \tau_f \rangle=\infty$,
diverges for one dimensional Brownian motion on the infinite  line. 
 $P(\chi_f|\tau_f)$ is the conditional
probability density function of $\chi_f$ given some fixed $\tau_f$. 
 This leads us to the concept of a Brownian excursion. 

\begin{figure}\begin{center}
\includegraphics[width=0.8\textwidth]{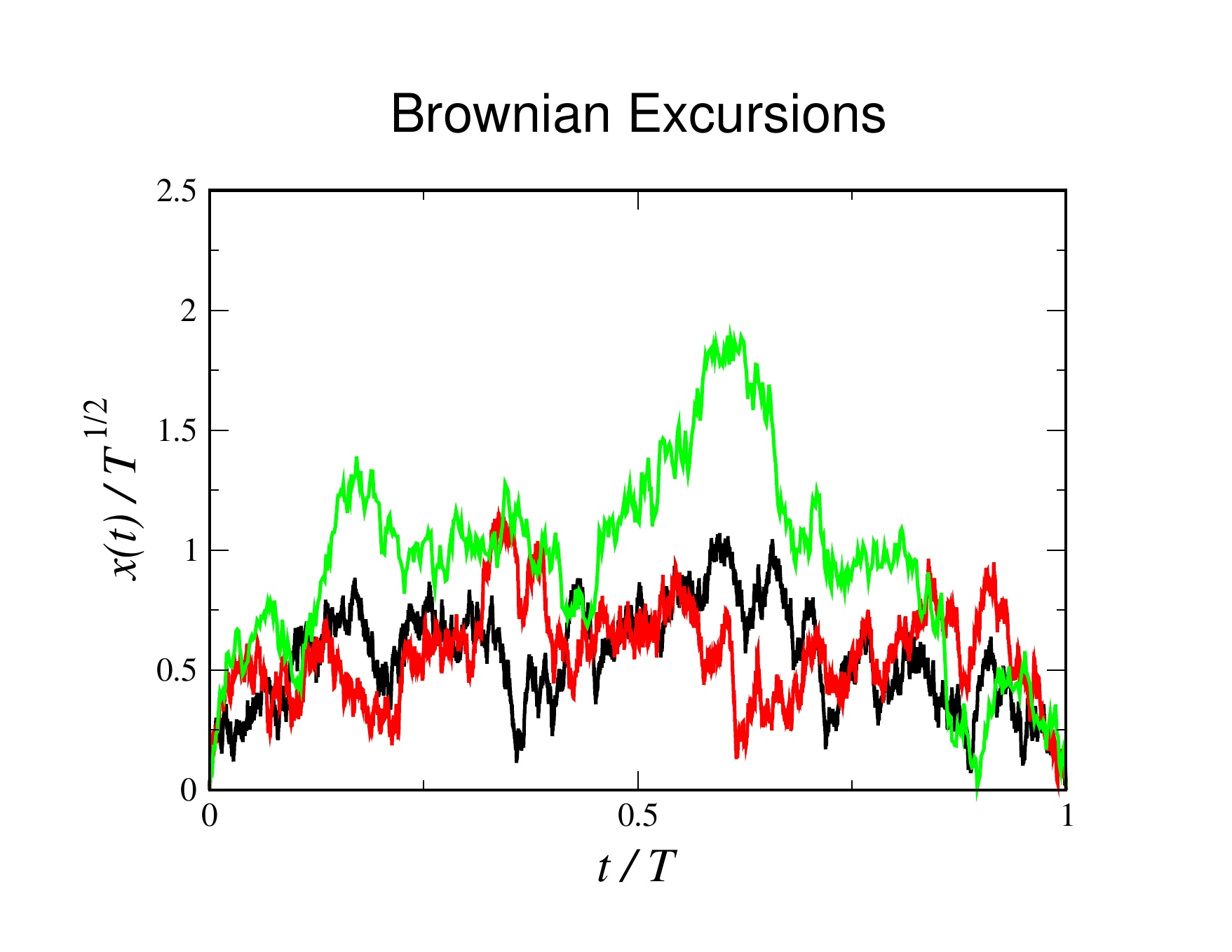}
\end{center}
\caption{
An example of a Brownian excursion $x(t)$, which in the text is generalized
to a Langevin excursion in momentum space, $p(t')$. These are random curves which  start and end
close to the origin and remain positive for a fixed time. Surprisingly
the random area under these curves has started to attract physical attention
in diverse problems ranging from fluctuating interfaces and as we show here diffusion of atoms in optical lattices. A related problem is the Tracy-Widom distribution. 
}
\label{figExcursion}
\end{figure}

 A Brownian excursion is a conditioned one dimensional
 Brownian motion  $x(t)$ over the time interval $0\le t \le \tau_f$ (recall
here $\tau_f$ is fixed to a specific value). The motion  starts
at $x(0)=\epsilon >0$, $\epsilon\ll 1$, and  ends at $x(\tau_f) = 0$ and is constrained not
to cross the origin $x=0$ in the observation time $(0,\tau_f)$, namely the
Brownian excursion is always positive. 
The area under the Brownian excursion is the object of interest
since that gives the conditional probability density $P(\chi_f|\tau_f)$.
Thus the statistics of the area under the Brownian excursion also
gives the joint PDF of the first passage time and the total
 area under the Brownian
motion. 

 The area under the Brownian excursion was treated by mathematicians 
with applications in computer science and graph theory and 
more recently  in the physical context of fluctuations of
interfaces \cite{Majumdar1,Majumdar2}.
 Majumdar and Comtet \cite{Majumdar2}
 give a concise introduction and derivation of the solution
to this problem in terms of the Airy distribution.
 We shall soon show that the solution of the dynamics
of the atomic packet demands the solution of a first passage time
in momentum space, and the distribution of the area under the random first
passage time curve. 
Thus the random area  under  a random curve terminated at its first passage
turns into a relevant problem, but, fortunately
or unfortunately, the apposite curve is not a simple Brownian one. 

So, for perspective, we first review what is known about the distribution of $A$, the random area under the Brownian excursion,
$A= \int_0 ^{\tau_f} x(t) {\rm d} t$ in the $\epsilon \to 0$ limit.
 Since Brownian motion
scales like $x\sim t^{1/2}$ the area $A$ scales like 
$\int_0 ^{\tau_f} x(t) {\rm d} t \propto (\tau_f)^{3/2}$. Thus
the distribution has the scaling behavior $P_{\tau_f} (A) = (\tau_f)^{-3/2} f(A/(\tau_f)^{3/2} )$, where the $(\tau_f)^{-3/2}$ prefactor stems from the normalization
condition.
 The Laplace $x \to u$ transform of $f(x)$,
$\hat{f}(u)$ was computed by Darling and Louchard \cite{Darling,Louchard} 
\begin{equation}
\hat{f}(u) = u \sqrt{ 2 \pi} \sum_{k=1} ^\infty \exp\left( - \alpha_k u^{2/3} 2^{-1/3}\right)
\end{equation}
(here, the diffusion constant of the Brownian motion, arising from the simple random walk used to model the motion, is $1/2$).
The $\alpha_k$'s are the magnitudes of the zeros of the Airy
function $Ai(z)$ on the negative real axis $\alpha_1=2.3381..$,
$\alpha_2 = 4.0879...$, $\alpha_3 = 5.5205..$ and for that reason the
distribution is called the Airy distribution.
Tak\'acs \cite{Takacs}  inverted the Laplace transform to find
\begin{equation}
f(x) = \frac{ 2 \sqrt{6} }{ x^{10/3} }\sum_{k=1} ^\infty e^{ - b_k /x^2} b_k ^{2/3} U( - 5/6, 4/3, b_k/x^2)
\label{eqAiry}
\end{equation}
where $b_k = 2 (\alpha_k)^3 / 27$ and $U(a,b,z)$ is the confluent hypergeometric
function \cite{Abr}. 
Moments and asymptotic behaviors of the Airy distribution 
can be found in Ref. \refcite{Majumdar2}.
The function can be calculated and plotted by truncating the sum at some $x$ dependent value. 

 From this, we can  deduce the power law exponent describing the tail of
the PDF of the area
swept under  the Brownian curve till its first passage event  $q(\chi_f)$, averaged over all $\tau_f$.
Clearly  the marginal PDF $q(\chi_f)$  is found from the joint PDF
 in the usual
way by integration
\begin{equation}
q(\chi_f) = \int_0 ^\infty g(\tau_f) P(\chi_f|\tau_f) {\rm d} \tau_f
\end{equation}
Using Eq. (\ref{eqfpS}) and the scaling behavior of the conditional 
PDF $P(\chi_f|\tau_f)$, 
we have
\begin{equation}
q\left( \chi_f \right) \propto \int_0 ^\infty (\tau_f)^{-3/2} (\tau_f)^{-3/2} f(\chi_f/(\tau_f)^{3/2}) {\rm d} \tau_f.
\end{equation}
Changing integration variables  to $y \equiv \tau_f/(\chi_f)^{2/3}$
we get  the asymptotic form
\begin{equation}
q(\chi_f) \propto (\chi_f)^{-4/3}. 
\label{eqKear}
\end{equation}
We see that the scaling exponent $3/2$ for $g(\tau)$ and $4/3$ for $q(\chi_f)$
are easy to obtain and are both related to the scaling of Brownian
motion $x \propto t^{1/2}$.
 The full distribution of $q(\chi_f)$ was recently found
by Kearney and Majumdar for finite $\epsilon$. We will
later deal with such a calculation in more rigor, but for now we just focus
on the exponents.   
We now turn to  showing how the generalization
of these concepts lead to the description of the dynamics of cold atoms.

\section{Model and Goal}

We wish to investigate the spatial
 density of the atoms,
$P(x,t)$.
The trajectory of a single particle is $x(t) = \int_0 ^t p(t) {\rm d} t/m$ where
$p(t)$ is its momentum.
Within the standard picture~\cite{CT,Zoller}
of Sisyphus cooling, two competing
mechanisms describe the dynamics.
The cooling force
\begin{equation}
F(p) = - \overline{\alpha} p /[1 + (p/p_c)^2 ]
\end{equation}
acts to restore the momentum to the minimum energy state $p=0$.
Momentum diffusion is governed by  a
diffusion coefficient which is momentum dependent,
$D(p) = D_1 + D_2 / [1 + (p/p_c)^2]$. The latter
describes momentum fluctuations which
lead to heating (due to random emission events).
We use dimensionless units, time $t \to
t \overline{\alpha}$, momentum $p \to p/p_c$,    the momentum diffusion
constant  $D=D_1/ (p_c)^2 \overline{\alpha}$ and
$x \to x m \overline{\alpha}  / p_c$.
For simplicity,
we set $D_2=0$  since
 it does not modify the asymptotic $|p|\to \infty$
 behavior of the
diffusive heating term, nor that of the force,
 and therefore  does not modify our main conclusions.
The Langevin equations
\begin{subequations}\label{eq05}
\begin{align}
\frac{{\rm d} p }{ {\rm d} t } &= F(p) + \sqrt{ 2 D} \xi(t), \label{eq05a}\\
\frac{{\rm d} x }{ {\rm d} t } &= p
\label{eq05b}
\end{align}
\end{subequations}
describe the dynamics in phase space.
Here the noise term is Gaussian, has zero mean
and is white $\langle \xi(t) \xi(t') \rangle = \delta(t- t')$.
The now dimensionless cooling force is
\begin{equation}
F(p) = - \frac{ p }{ 1 + p^2} .
\label{eq03}
\end{equation}
The stochastic Eq. (\ref{eq05})
give the trajectories of the standard
Kramers picture  for   the semi-classical dynamics in the optical lattice
which in turn was derived from microscopical considerations \cite{CT,Zoller}.
 From the semiclassical treatment of the interaction of the atoms with
the counterpropagating laser beams, we have
\begin{equation}
D= c E_R/ U_0,
\end{equation}
where $U_0$ is the depth of the optical potential and  $E_R$
the recoil energy, and the dimensionless parameter
$c$
depends on the atomic transition involved.\footnote{For atoms in molasses
with a $J_g=1/2 \rightarrow J_e =3/2$
Zeeman substructure in a lin $\bot$ lin laser configuration $c=12.3$
\cite{Zoller}.
Refs. \refcite{CT,Lutz,Renzoni}
give  $c=22$.
As pointed out  in \cite{Zoller} different notations are used in
the literature.}
 For $p \ll 1$, the cooling force Eq. (\ref{eq03})
is  harmonic, $F(p) \sim - p$,
while in the opposite limit, $p\gg 1$, $F(p) \sim - 1/p$. It is useful, given the correspondence between Eq.
(\ref{eq05a}) and overdamped motion in space subject to a force $F$, to define an effective
potential 
\begin{equation}
V(p) = - \int_0 ^p F(p) {\rm d} p= (1/2) \ln(1 + p^2) .
\end{equation}
This potential is asymptotically logarithmic, $V(p) \sim \ln(p)$
when $p$ is large, which has as its consequence
various unusual equilibrium and non-equilibrium
properties of the momentum distribution
\cite{Katori,Lutz,Renzoni,KesslerPRL,KesslerJStatPhys,Mukamel2}.
For example the steady state momentum distribution
is $W_{eq}(p) \propto \exp[ - V(p)/D]$ which has a power law form
\begin{equation}
W_{eq}(p) = {\cal N} (1 + p^2)^{ -1 / 2 D}  
\end{equation}
where ${\cal N}$ is the normalization. 
Such non-Gaussian behavior
was observed in the experiments of Renzoni's group~\cite{Renzoni}.
For $D< 1/3$, corresponding to shallow optical lattices, the averaged
kinetic energy diverges. A similar effect was found experimentally
by Walther's group~\cite{Katori}: when the depth of the optical lattice is tuned
one may observe a gigantic increase of the energy of the system.
Of course the energy of a physical system cannot be infinite and this paradox
 was recently resolved~\cite{KesslerPRL,KesslerJStatPhys} by treating the problem dynamically, revealing that
the energy may increase with time but it is never infinite. Of course,
when the energy becomes too large, other experimental effects become relevant as well.  

While the velocity distribution of the atomic cloud
appears to be well understood, the new experiment \cite{Sagi}
demands a theory
for the spatial spreading. This will be the focus of this chapter.
However, by now the reader might think that laser cooling is an
oxymoron. Namely if we get non-Maxwellian distributions in equilibrium,
fat tails, and L\'evy behavior, where is the thermal
state found? The answer, as we understand it, is rather involved.
 First the equilibrium momentum PDF
can be rather narrow in its center, and in that sense the atoms are
cold. Further, the divergence of energy happens for low $D$,
so an experimentalist interested in cooling will turn their knobs away
from that limit. In fact there is an optimal depth of the optical
potential in the sense that it yields the smallest value for  the mean kinetic energy of the particles, 
since the basic energy scale is set by the optical potential depth
(see, e.g., Fig. 4 of Ref. \refcite{Katori} and Fig. 2 of Ref. 
\refcite{DechantPNAS}).
 Strictly speaking, however,  for all $D$ the
equilibrium state is not thermal since the momentum distribution is not
Gaussian.
Experimentalist intuitively recognize this problem, and hence also use 
a method called evaporation to reach an ultimate cold state. Evaporation
simply  means that atoms in the tails, i.e. atoms with high velocities, are
a problem for cooling, so they are allowed  by the experimentalist to leave the system.
 Another mechanism that
may lead to a realistic thermal state is internal collisions
among the atoms (which will depend of-course on the density).
It is believed that such  effects will
lead the system in the long time limit to thermal equilibrium.
However, at least in the experiments briefly
surveyed here~\cite{Katori,Renzoni,Sagi}, which take long times
on atomic scale (e.g. milliseconds) no hint of thermal Maxwellian
equilibrium is found, at least for values of $D$ of interest. Hence
it is an important task to map the different phases of the
dynamics which are controlled by $D$. Indeed this system is truly
wonderful in the sense that the experimentalist can tune $D$ and hence
explore different  effects both in and out of equilibrium. 

\begin{figure}\begin{center}
\includegraphics[width=0.8\textwidth]{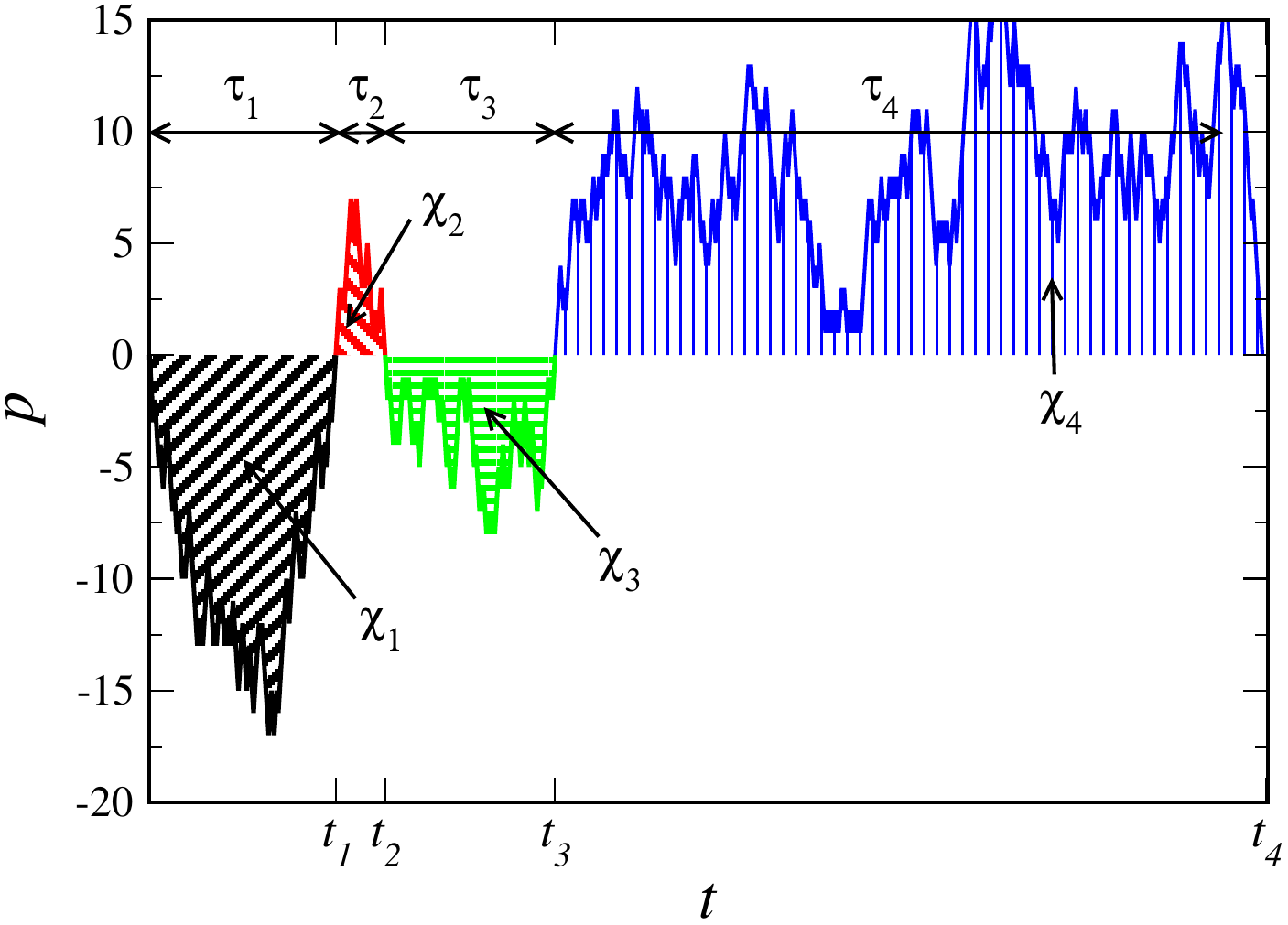}
\end{center}
\caption{
Schematic presentation of the momentum of the particle versus time.
The times between consecutive
zero crossings are called the jump durations $\tau$ and the shaded
area under each excursion are the random flight displacements $\chi$.
The $\tau$'s and the $\chi$'s are correlated, since statistically
a long jump duration implies a large displacement.
}
\label{fig1}
\end{figure}

\section{From Area under the Bessel excursion to L\'evy walks for
the packet of atoms}

 The heart of our analysis is the mapping of the Langevin dynamics
 to a recurrent set of random walks.
 The particle along its stochastic path in momentum space
crosses $p=0$ many times when the measurement time is long.
Let $\tau>0$ be the random time between one crossing event to
the next crossing event,
and let $-\infty <  \chi  < \infty$ be the
random displacement (for the corresponding $\tau$).
 As schematically shown in Fig. \ref{fig1},
the process starting at the origin with
zero momentum is defined by the sequence of jump durations, $\left\{ \tau_1, \tau_2 , \ldots\right\}$ with corresponding
displacements $\left\{ \chi_1,  \chi_2, \ldots \right\}$, with
$\chi_1 \equiv \int_0^{\tau_1} p(\tau) d\tau$, $\chi_2 \equiv \int_{\tau_1}^{\tau_1+\tau_2} p(\tau)d\tau$, etc.
 The total
displacement $x$ at time $t$ is a sum of the individual
displacements $\chi_i$.
 Since the underlying Langevin dynamics
is continuous,
we need a more precise definition of this process.
Let $\tau$ be  the time it takes the particle
with initial momentum $p_i$ to reach $p_f=0$ for the
first time. So $\tau$ is the first passage time for the process
in momentum space.  Eventually we will take  $p_i \to p_f$, see below.
Similarly, $\chi$ is the displacement of the particle
during this flight. Namely by definition
$\chi$ is the area under the 
Langevin curve (in momentum space) till the first passage time.  
The probability density function (PDF) of the displacement
$\chi$ is denoted $q(\chi)$ and
of the jump durations $g(\tau)$.
To conclude we see that first passage time statistics, in particular
 the area swept under
the random momentum process $p(\tau)$ until the first passage time, 
constitute the microscopical random jumps 
and correlated random waiting times
which eventually give the full displacement of the the particle.

As shown by Marksteiner, et al. and Lutz  \cite{Zoller,Lutz}, these PDFs
exhibit power law behavior
\begin{subequations}\label{eq06}
\begin{align}
g\left( \tau \right) &\propto \tau^{-\frac{ 3}{ 2} - \frac{1}{2 D}},\label{eq06a}\\
  q\left(\chi\right) &\propto | \chi|^{ - \frac{4 }{3}  -\frac{1 }{ 3 D} }, \label{eq06b}
\end{align}
\end{subequations}
as a consequence of the logarithmic potential,
which makes the diffusion for large enough $p$ only weakly bounded.  It is this power-law behavior, with its divergent second moment of the displacement $\chi$ for $D>1/5$, which gives rise to
the anomalous statistics for $x$ (measured in the experiment).
Notice that taking the limit $D\to \infty$ in Eq. (\ref{eq06})
 we get the expected results as given in
Eqs. (\ref{eqfpS}) and (\ref{eqKear}).
In this limit, the cooling force $F(p)$ is negligible, the dynamics
is governed by the noise in Eq. (\ref{eq05})
and so the process $p(t)$ approaches a Brownian motion. 
This corresponds to atoms emitting a photon in random directions,
and hence from the recoil jolts, they are undergoing an unbiased
 random walk in momentum
space (which is the pure heating limit). 

Importantly, and previously overlooked, there is a
strong correlation between the jump duration $\tau$ and
the spatial extent of the jumps $\chi$.
These correlations are responsible,
in particular, for the finiteness of the moments of $P(x,t)$.
Physically, such a correlation is obvious, since long
jump durations involve large momenta, which in turn
induce a large spatial displacement.
 The theoretical development starts then from the quantity
 $\psi( \chi, \tau)$,
the joint probability density of  $ |\chi|$
and $\tau$.
From this, we construct a L\'evy walk
scheme \cite{KBS,Blumen,Carry} which relates the microscopic
information $\psi(\chi,\tau)$ to the atomic packet $P(x,t)$ for large $x$ and $t$.

\section{Scaling Theory for Anomalous Diffusion}
\begin{figure}[h]\begin{center}
\includegraphics[width=0.7\textwidth]{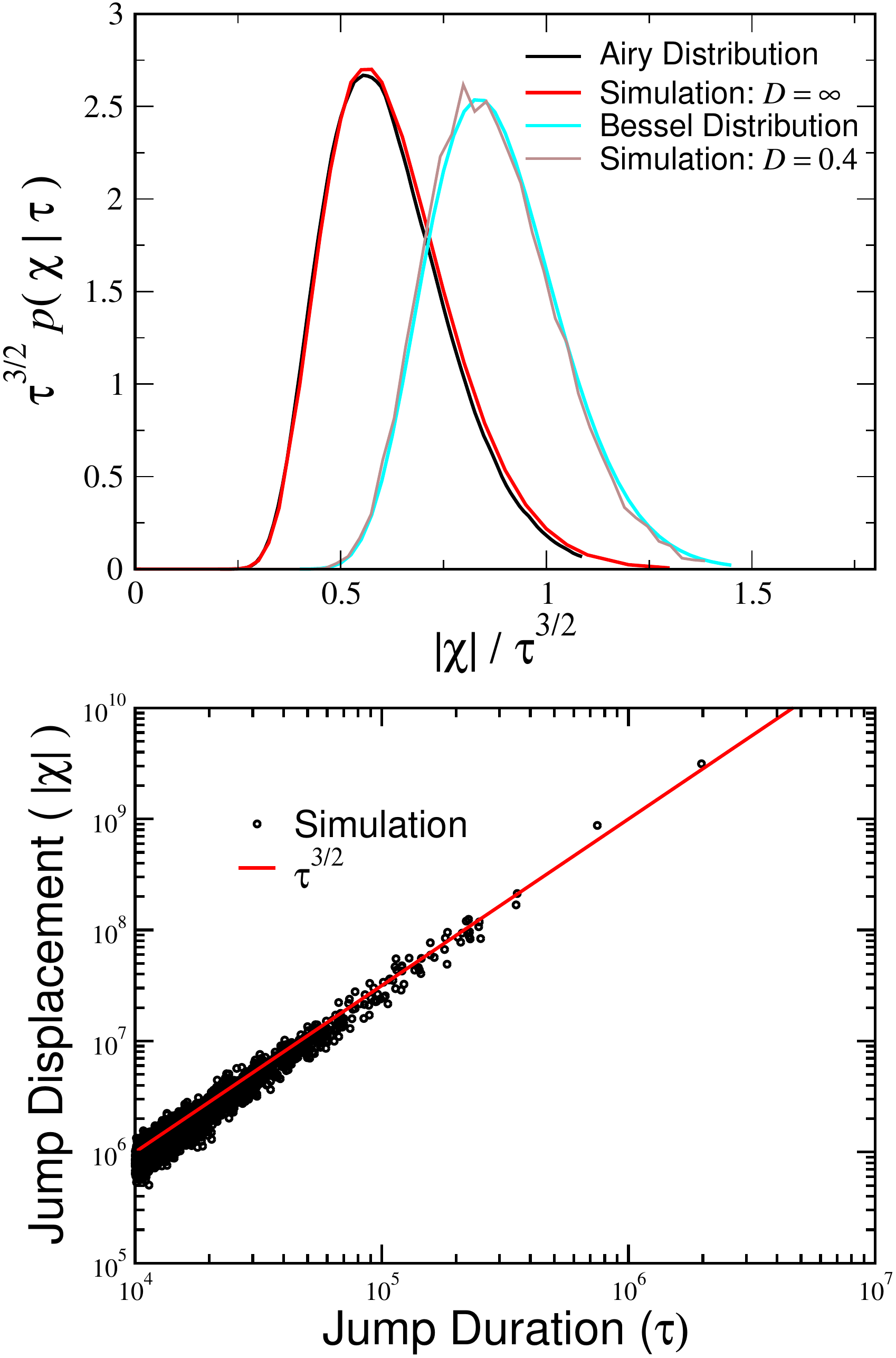}
\end{center}
\caption{
Lower panel:  We plot the flight displacement $|\chi|$ versus
the jump duration $\tau$ to demonstrate the strong correlations between
these two random variables. Here $D=2/3$ and the  line has  slope $3/2$
reflecting the $\chi\sim \tau^{3/2}$ scaling discussed in the text.
Upper panel: The conditional probability $p(\chi|\tau)$ is
described by the Airy distribution~\cite{Majumdar1,Majumdar2} when $D \to \infty$; otherwise
by the distribution derived here based on the Feynman-Kac theory,
 Eq.
(\ref{eqHG}) .
They describe
areas under constrained Brownian (or Bessel) excursions
 when the cooling force $F(p)=0$ (or $F(p) \ne 0$)
respectively.
}
\label{fig2}
\end{figure}

We rewrite the joint PDF
\begin{equation}
\psi( \chi, \tau)=\frac{1}{2}g(\tau) p\left(| \chi|\, \big{|}\,\tau\right),
\label{joint}
\end{equation}
where $p(\chi \, | \,\tau)$ is the conditional probability to find a jump length of
$ 0<\chi <\infty$ for a given jump duration $\tau$. The factor $\tfrac{1}{2}$ accounts for the walks with $\chi<0$.
Numerically we observed that the conditional probability scales
at large times like
\begin{equation}
p(\chi |\tau) \approx 2\tau^{- \gamma} B(|\chi| / \tau^\gamma); \qquad\qquad \gamma=3/2.
\label{eqSca}
\end{equation}
$B(\cdot)$ is a scaling function soon to be determined.
 To analytically obtain  the scaling exponent $\gamma=3/2$ note
that
$q\left(\chi\right)=\int_0 ^\infty {\rm d} \tau \psi \left( \chi,\tau \right)$, giving
\begin{equation}
q \left(  \chi\right) \sim
 \int_{\tau_0}  ^\infty {\rm d} \tau \tau^{-\frac{3}{2 }- \frac{1 }{2 D}}  \tau^{- \gamma} B \left(\frac{| \chi | }{ \tau^\gamma}\right)
 \propto | \chi|^{ - \left(1 + \frac{1+1/D}{2\gamma}\right)} .
\label{eq07}
\end{equation}
Here $\tau_0$  is a time scale after which the long time
limit in Eqs. (\ref{eq06}) holds and is irrelevant for large $\chi$.
Comparing Eq. (\ref{eq07}) to   Eq.
(\ref{eq06b})  yields
   the consistency condition
 $1 + (1 + 1/D)/(2 \gamma) =
4/3 + 1/ (3 D)$ and hence $\gamma=3/2$, as we observe in Fig. \ref{fig2}.

We see that the scaling function $B(.)$ in Eq. (\ref{eqSca}) 
is similar to the problem of the area under a Brownian excursion. 
 However, now we are
confronted with the additional impact of the friction force. 
The concept of the Brownian excursion needs to be extended to a stochastic curve called
the Bessel excursion~\cite{Hu}, which corresponds to with the friction force $F(p)=-1/p$. 
The term Bessel excursion stems from the fact that mathematically speaking,
 diffusion in momentum space, in the non-regularized potential $\ln(p)$,
corresponds to a process called the Bessel process~\cite{Bray,Schehr}.
As with the Brownian excursion, the Bessel excursion  is the Langevin path $p(t')$
over the time interval $0\le t' \le \tau$ such that
the path starts on $p_i \to 0$ and ends on the
origin, but is {\em constrained} to stay positive (if $p_i>0$) or
negative (if $p_i<0$). Brownian excursions are just the $D\to\infty$
 limit of this more general problem.
  Schematic excursions $p(t)$ are presented in Fig.
\ref{fig1}, albeit there with various excursion times $\tau$.
The displacement $\chi= \int_0 ^\tau p(t') {\rm d} t'$ is
the area under the  excursion as shown in Fig. \ref{fig1}.

\subsection{Area Under the Bessel Excursion}

We briefly outline how to obtain
the distribution of the area under the Bessel excursion, namely we
find the rather formidable $p(\chi|\tau)$, Eq.
(\ref{eqHG}) below.
The main tool we use is a modified Feynman-Kac theory. The original
well-known formalism deals with functionals of Brownian motion,
while here we deal with functionals of Langevin motion, hence the modification.
The Feynman-Kac equation is essentially a Shr\"odinger equation in real
time and is reviewed in Ref. \refcite{MajRev}. 
 The calculation involves three steps:
\begin{itemize}
 \item[(a)] We first find $\widetilde{G}(s,p,\tau)$, the Laplace transform with respect to $\chi$
of the joint PDF  for the positively constrained path to reach  $p$ with area $\chi$ at time $\tau$, which we denote by
$G(\chi, p,t)$.\\
\item[(b)]
Taking the limit of  the initial and final momentum $p_i=p \to 0$,
we get
 $\tilde{p}(s|\tau)=\lim_{p,p_i\to 0} \widetilde{G}(s,p,\tau)/\widetilde{G}(s=0,p,\tau);$
the $\tilde{G}(s=0,p, \tau)$ denominator yields the correct normalization.
\\ 
\item[(c)] Finally, using the inverse Laplace transform, we get $p(\chi\,\big{|}\,t)$.
\end{itemize}
As mentioned, the main tool we use is the modified Feynman-Kac formalism for functionals
of over-damped Langevin paths \cite{MajRev,Carmi}. For step (a) we solve:
\begin{equation}
\frac{\partial \widetilde{G}(s,p,\tau) }{ \partial \tau} = \left[ \widetilde{L}_{{\rm fp}} - s p \right] \widetilde{G}(s,p,\tau),
\label{eqFK}
\end{equation}
where $\widetilde{L}_{{\rm fp}} = D (\partial_p)^2 - \partial_p F(p) $ is the Fokker Planck operator. When $F(p)=0$, Eq. (\ref{eqFK}) is
the celebrated Feynman-Kac equation which is an imaginary-time Schr\"odinger
equation in a linear potential (i.e. the $-s p$ term).
We do not provide here a derivation of Eq. (\ref{eq21}) but in
Sec. \ref{Sec08} provide a derivation of a related equation.  The
constraint $p>0$ gives the boundary condition
$\widetilde{G}(s,p=0,t)=0$, i.e., an infinite potential barrier
when $p<0$ in the quantum language.
We use the force $F(p) = - 1/p$ since we are
interested only in the scaling behavior of the problem where large
excursions imply that the small $p$ behavior of the force is not important.
This assertion can be mathematically justified,
but here for the sake of space  we will only demonstrate it numerically
(see Fig. \ref{fig2}).
  Solving Eq. (\ref{eqFK}) and following the recipe
(a) - (b) we find
\begin{equation}
\widetilde{p}(s|\tau) = \sum_{k=1} ^\infty 2^{3 \nu + 1} \Gamma\left( 1 + \frac{3 }{ 2} \nu\right) \left( s D^{1/2} \tau^{2/3} \right)^{\nu +2/3} [g'_k(0)]^2 e^{ - \tau D^{1/3} \lambda_k s^{2/3} } .
\end{equation}
Here $D  g_k'' +  (g_k/p)' - p g_k = - D \lambda_k g_k$
is the eigenvalue equation which gives the eigenvalues $\lambda_k$ and
eigenstates $g_k(p)$ which are normalized according
$\int_0 ^\infty [g_k(p)]^2 p^{1/D} {\rm d} p =1$.
A detailed mathematical derivation will soon be published \cite{KesBarPRX}.

It is easy to  tabulate the eigenvalues
$\lambda_k$ and the slopes on the origin $g'_k(p=0)$
with standard numerically exact techniques.
With this information
and the inverse Laplace transform we find the
solution in terms of generalized hypergeometric functions
\begin{eqnarray} p(\chi\,\big{|}\,\tau) &=&
  -\frac{\Gamma\left(1+\frac{3\nu}{2}\right)}{4\pi \chi} \left(\frac{4D^{1/3}\tau}{\chi^{2/3}}\right)^{1+3\nu/2}\times\hspace*{\fill}  \nonumber\\
  &\ &
\sum_k [g'_k(0)]^{2}\left[  {}_2F_2\left(\frac{4}{3}+\frac{\nu}{2},\frac{5}{6}+
\frac{\nu}{2};\frac{1}{3},\frac{2}{3};-\frac{4D\lambda_k^3\tau^3}{27 \chi^2}\right) c_0
\right.\nonumber\\
&\ &\qquad\qquad {} -  {}_2F_2\left(\frac{7}{6}+\frac{\nu}{2},\frac{5}{3}+\frac{\nu}{2};\frac{2}{3},\frac{4}{3};-\frac{4D\lambda
_k^3\tau^3}{27 \chi^2}\right)c_1 \nonumber\\
&\ &\qquad\qquad \left. {} + {}_2F_2\left(2
+\frac{\nu}{2},\frac{3}{2}+\frac{\nu}{2};\frac{4}{3},\frac{5}{3};-\frac{4D\lambda_k^3\tau^3}{27 \chi^2}\right)\frac{c_2}{2}\right],\nonumber\\
&\ &
\label{eqHG}
\end{eqnarray}
where 
\begin{equation}
c_j = \left(\frac{D^{1/3}\lambda_k \tau}{
\chi^{2/3}}\right)^j \Gamma\left(\frac{5+2j}{3}+\nu\right)\sin\left(\pi\frac{2 + 2j+3\nu}{3}\right) \qquad j=0\ldots 2
\end{equation}
and the summation is over the eigenvalues.
Here we have introduced the parameter
\begin{equation}
\nu \equiv \frac{1+D}{3D}\ 
\end{equation}
which will turn out to play a crucial role in the dynamics. 
Notice the $\chi \sim \tau^{3/2}$ scaling which proves the
scaling hypothesis, Eq.
(\ref{eqSca}).
To compare with numerical simulations, we evaluate this function truncating the sum beyond some $k$, since for fixed $x$ the terms decay rapidly in $k$. The Langevin simulations  were performed by mapping the
problem onto a
biased random walk in momentum space,
 with momentum spacing $\delta p = 0.1$.

In Fig. \ref{fig2}, we present $p(\chi|\tau)$
obtained from numerical simulations, showing that it perfectly matches our
theory on Bessel excursions without fitting.
Our theory and simulations show that
the areas under the Brownian and Bessel excursions share the same
$\chi \sim \tau^{3/2}$ scaling behavior.
In both cases, $p(\chi|\tau)$ falls off rapidly for large $|\chi|$, ensuring
finite moments of this distribution, which will prove important later.
When $D\gg 1$, the Bessel and the Brownian excursions
coincide, since then the force $F(p)$ plays a vanishing role,
as can be shown directly from Eq. (\ref{eqHG}).  The Bessel excursion PDF for finite $D$ is compared with the limiting Airy distribution in Fig. \ref{fig2},
where it is seen that the Bessel excursions
are pushed further away from the origin.
The  $-1/p$ force implies that trajectories not crossing
the origin in $(0,\tau)$ are typically further away from the origin,
compared  with the free Brownian particle, since
trajectories close to the origin are more likely to cross it,
due to the attractive force.

\subsection{Coupled Continuous Time Random Walk Theory from First Passage
Time Statistics}

The next step is to
relate  the 
first passage time properties of the underlying Langevin process 
in momentum space, namely the joint distribution
of the area under the Bessel excursion and the first passage time,
with the spreading of the density  $P(x,t)$. 
Given our scaling solution for
$p(\chi|t)$, and $g(\tau)$ which is considered in detail
in Ref. \refcite{KesBarPRX} (see also Appendix here)
 and hence $\psi(\chi,t)$, 
we construct a theory for $P(x,t)$
using tools developed in the random walk community
\cite{KBS}.  As soon detailed, 
one first obtains a Montroll-Weiss \cite{Review} type of equation for the Fourier-Laplace transform of $P(x,t)$, $\widetilde{P}(k,u)$, in terms of $\widetilde{\psi}(k,u)$,
the Fourier-Laplace transform of the joint PDF $\psi(\chi,\tau)$:
\begin{equation}
\widetilde{P}(k,u) = \frac{\Psi(k,u)}{1-\widetilde{\psi}(k,u)} .
\label{MW}
\end{equation}
Here, $\Psi(k,u)$ is the Fourier-Laplace transform of $\tau^{-3/2}B(|\chi|/\tau^{3/2})[1-\int_0^t \psi(\tau) d\tau]$. The last step is then to invert Eq. (\ref{MW})
back to the $x$, $t$ domain.

 To derive Eq. (\ref{MW})
we present a formalism which relates the joint distribution
\begin{equation}
\psi( \chi, \tau) = g(\tau) \tau^{-3/2} B\left( | \chi| / \tau^{3/2}\right)
\end{equation}
 with $P(x,t)$ ($B(.)$ is given in Eqs. 
(\ref{eqSca},\ref{eqHG})). 
Define $\eta_s(x,t) {\rm d} t {\rm d} x$ as the probability that the
particle crossed the  momentum state $p=0$ for the $s^\textit{th}$ time in
the time interval $(t,t+{\rm d} t)$ and that the particle's position
was $(x,x + {\rm d} x)$.
This probability is related to the probability of the  $s-1$ crossing
according to
\begin{equation}
\eta_s(x,t) = \int_{-\infty} ^\infty {\rm d} v_{3/2} \int_0 ^\infty {\rm d} \tau\,  \eta_{s-1}\!\!\left( x - v_{3/2} \tau^{3/2} , t- \tau\right) B\left(|v_{3/2}| \right)
g\left( \tau \right) 
\label{eq09}
\end{equation}
where we changed the variable of integration to $v_{3/2} \equiv \chi/\tau^{3/2}$. 
Now the process is described by a sequence of jump durations
$\tau_1, \tau_2, \cdots$ and the corresponding generalized velocities
$v_{3/2}(1), v_{3/2}(2) , \cdots$
(with units $[x]/[t^{3/2}]$).
The displacements in the $s$th
  interval is:  $ \chi_s = v_{3/2}(s) [\tau_s]^{3/2}$.
The advantage of the representation of the problem
in terms of the pair of  microscopic stochastic variables
$\tau,v_{3/2}$
(instead of the correlated pair $\tau,  \chi$)
is clear from Eq. (\ref{eq09}): we may treat $v_{3/2}$ and $\tau$ as
independent random variables whose corresponding PDFs are
$g(\tau)$ and $B(v_{3/2})$ respectively.
The initial condition $x=0$ at time $t=0$
implies $\eta_0(x,t) = \delta(x) \delta(t)$.
Let $P(x,t)$ be the probability of finding the particle in
$(x,x+ {\rm d} x)$ at time $t$, which is found according to
\begin{equation}
P(x,t) = \sum_{s=0} ^\infty \int_{-\infty} ^\infty {\rm d} v_{3/2} \int_0 ^\infty {\rm d} \tau\, \eta_s\!\!\left( x - v_{3/2} \tau^{3/2} , t - \tau \right) B\left(|v_{3/2}| \right) W \left( \tau \right).
\label{eq10}
\end{equation}
since for a time series with $s$ intervals,
the last jump event took place at $t - \tau$ and in the time period
$(t-\tau,t)$ the particle did not cross the momentum origin (hence the
survival probability $W(\tau)=1 - \int_0 ^\tau g(\tau) {\rm d} \tau$).
 The summation in Eq. (\ref{eq10}) is
a sum of all possible realizations with $s$ returns to the momentum
origin $p_f =0$.

As usual \cite{Review,KBS},
we consider the problem in Laplace-Fourier space where
 $t \to u$ and $x \to k$. Eq. (\ref{eq09}) then translates to
\begin{equation}
\tilde{\eta}_s\!\!\left( k , u \right) = \tilde{\eta}_{s-1}\!\!\left( k , u \right) \widetilde{\psi}(k,u).
\label{eq11}
\end{equation}
Here the tildes refer to the Fourier-Laplace transforms, and we have used Eq. (\ref{eqSca}) to write that $\psi(\chi,\tau)=g(\tau)\tau^{-3/2}B(|\chi|/\tau^{3/2})$.
 Hence, summing
the Fourier-Laplace transform of Eq. (\ref{eq10})  using Eq.
(\ref{eq11}),  we get  the Montroll-Weiss  Eq. (\ref{MW}).

 The Montroll-Weiss equation is widely used in applications of 
the random walk. This equation holds since the underlying dynamics
is Markovian, namely each time the process $p(t)$ crosses  zero,
the process is renewed. However, there is a subtle point behind our
approach, which makes it different. Since the trajectories are continuous, the number of 
zero crossings $s$ is infinite within any finite time interval
$(0,t)$, and for a particle starting with $p=0$. In contrast
in usual continuous time random walk theories the number of renewals
is always finite, and it increases with time. The functions entering
the right hand side of  Eq. 
(\ref{MW}) still depend on the initial momentum we assign to the particle 
$\pm p_i$ immediately after each zero crossing (if we use a grid in a numerical simulation this
is handled automatically).  We still must take the limit $p_i\to 0$ and show
that our final expression will give sensible results in that regularized
limit (see also Ref. \refcite{KesBarPRX}).

\section{The L\'evy phase}

 We now explain why L\'evy statistics describes the diffusion profile $P(x,t)$ when $1/5 < D < 1$, provided that $x$ is not too large.
The key idea is  that, for $x$'s which are large,
 but not extremely large, the problem decouples, and $\widetilde{\psi}(k,u)$ can be expressed as a product of the
Fourier transform of $q(\chi)$, $\tilde{q}(k)$ and the Laplace transform 
of $g(\tau)$, $\tilde{g}(u)$.  This is valid as long as $x \ll t^{3/2}$,
 since otherwise
paths where $\chi \sim t^{3/2}$ are relevant, for which the correlations 
are strong, as we have seen.  The long-time, large-$x$
 behavior of $P(x,t)$ in the
decoupled regime is then governed by
the small-$k$ behavior of  $\tilde{q}(k)$
 and the small $u$ behavior of  $\tilde{g}(u)$. 
   When the second moment of $q(\chi)$ diverges, i.e. for $D>1/5$, 
the small-$k$ behavior  of $\tilde{q}(k)$ is determined by the 
large-$\chi$ asymptotics of $q(\chi)$ as given in Eq. (\ref{eqSca}),
$q(\chi) \sim  x_*^\nu/|\chi|^{1+\nu}$. 
When the first moment of $\tau$ is finite, i.e. for $D<1$, 
the small-$u$ behavior of  $\tilde{g}(u)$ is
governed by the first moment, $\langle \tau \rangle$.  From these follow the small-$k$, small-$u$ behavior of $\widetilde{P}(k,u)$:
\begin{equation}
\widetilde{P}(k,u) \sim \frac{1}{ u +  K_\nu |k|^\nu}
\label{Levyku}
\end{equation}
where $K_\nu = \pi x_*^\nu / (\langle \tau \rangle \Gamma(1+\nu)\sin \frac{\pi\nu}{2})$
(see some details in Appendix A).

 Both $x_*^\nu$ and $\langle \tau\rangle$ can be calculated (see Appendix A and Ref.
\refcite{KesBarPRX})
 via appropriate backward Fokker-Planck equations.  They both vanish as the magnitude of the initial momentum of the walk goes to zero, but their ratio has a finite limit, so that $K_\nu$, upon returning to dimensionfull units, is
\begin{equation}
K_\nu = \frac{ \sqrt{\pi} (3 \nu -1)^{\nu -1} \Gamma\left(\frac{3\nu-1}{2}\right) }{ \Gamma\left(\frac{3\nu-2}{2}\right) 3^{2 \nu-1} [\Gamma(\nu)]^2 \sin\left( \frac{\pi \nu}{2} \right)}
\left(\frac{p_c }{ m} \right)^\nu ( \overline{\alpha})^{- \nu + 1}.
\label{eq21}
\end{equation}
$\widetilde{P}(k,u)$, as given in Eq. (\ref{Levyku}), is in fact precisely the symmetric L\'evy distribution in Laplace-Fourier space with index $\nu$, whose $(x,t)$ representation is (see Eq. (B17) of Ref. \refcite{Bouchaud})
\begin{equation}
P(x,t) \sim \frac{1 }{\left(K_\nu t\right)^{1/\nu}}   L_{\nu,0}\left[\frac{x }{
\left( K_\nu  t^{1/\nu}\right)} \right].
\label{eq18}
\end{equation}
The properties of the L\'evy function $L_{\nu,0}(.)$ are well-known, and
the solution in time for the spatial Fourier transform is $P(k,t) = \exp( - K_\nu t |k|^\nu)$
with $2/3<\nu<2$. 
This distribution is the solution of
 the fractional diffusion equation, Eq. (\ref{eq01}), with $\beta=1$ 
and an initial distribution located at the origin.  This justifies 
the use of Eq. (\ref{eq01}) in Ref. \refcite{Sagi} for $1/5 < D < 1$
 and provides $\nu$ and $K_\nu$ in terms of the experimental parameters. 
 We can verify this behavior in simulations, as shown in Fig. 3, where we
see excellent agreement to our theoretical prediction,
 Eqs. (\ref{eq18}) and (\ref{eq21}), without any fitting.

 The fractional diffusion equation, Eq. \ref{eq01},
 has the following meaning and
connection to the underlying random walk.
We re-express Eq. (\ref{Levyku}) as $u\widetilde{P}(k,u) - 1 =- K_\nu |k|^{\nu}\widetilde{P}(k,u)$.  Recalling~\cite{Review} that the Fourier
space representation of the Weyl-Rietz fractional derivative, 
$\nabla^\nu$, is $-|k|^\nu$ and that $u\widetilde{P}(k,u) -1$
 is the representation of the time derivative in Laplace space for 
$\delta$-function initial conditions centered at the origin, we see that in fact $P(x,t)$ satisfies the fractional diffusion equation, Eq. (\ref{eq01}),
with $\beta=1$.

\begin{figure}[ht]
\begin{center}
  \subfigure[]
     {\epsfig{figure=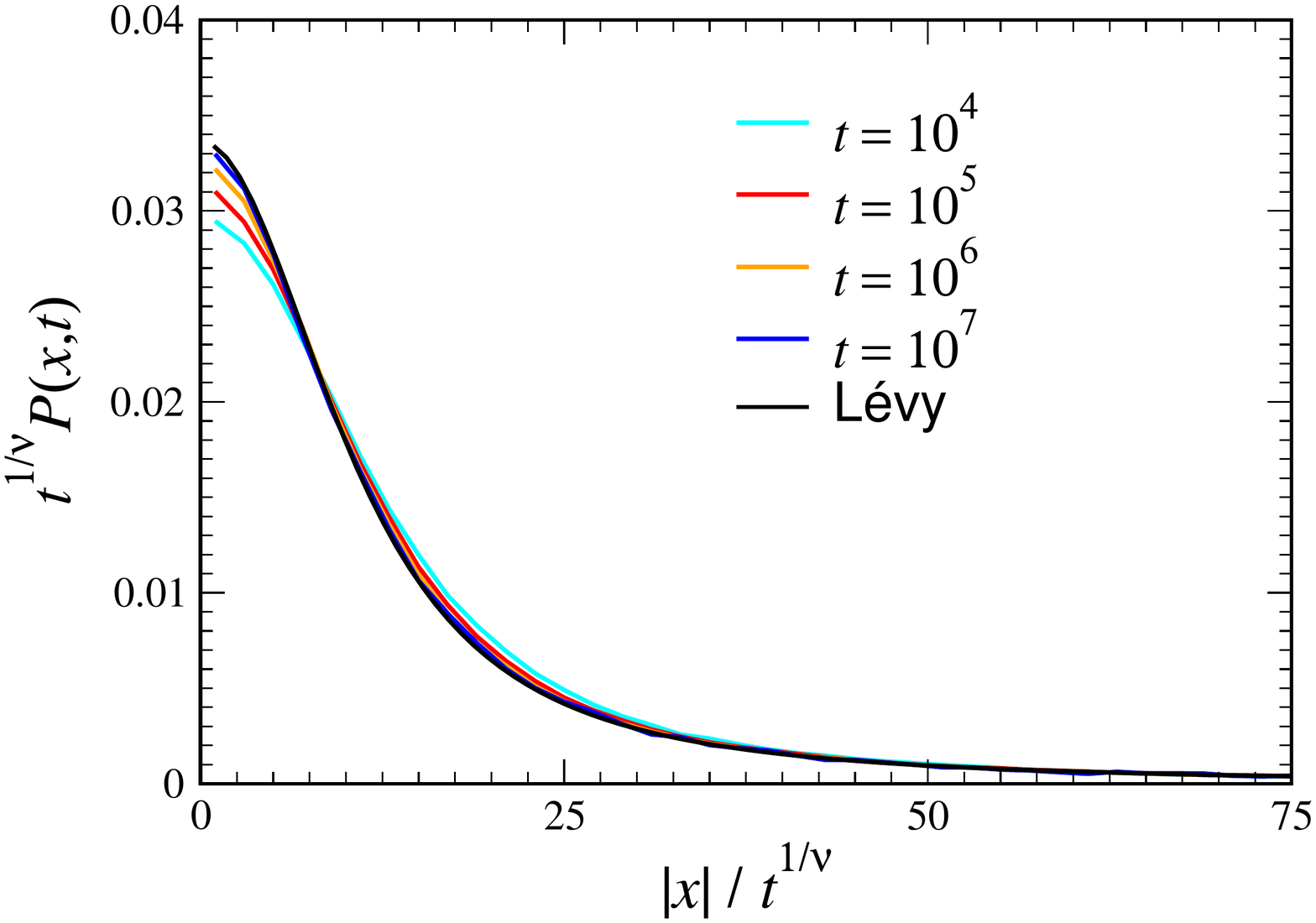,width=2.1in}\label{ra_fig2a}}
  \hspace*{4pt}
  \subfigure[]
     {\epsfig{figure=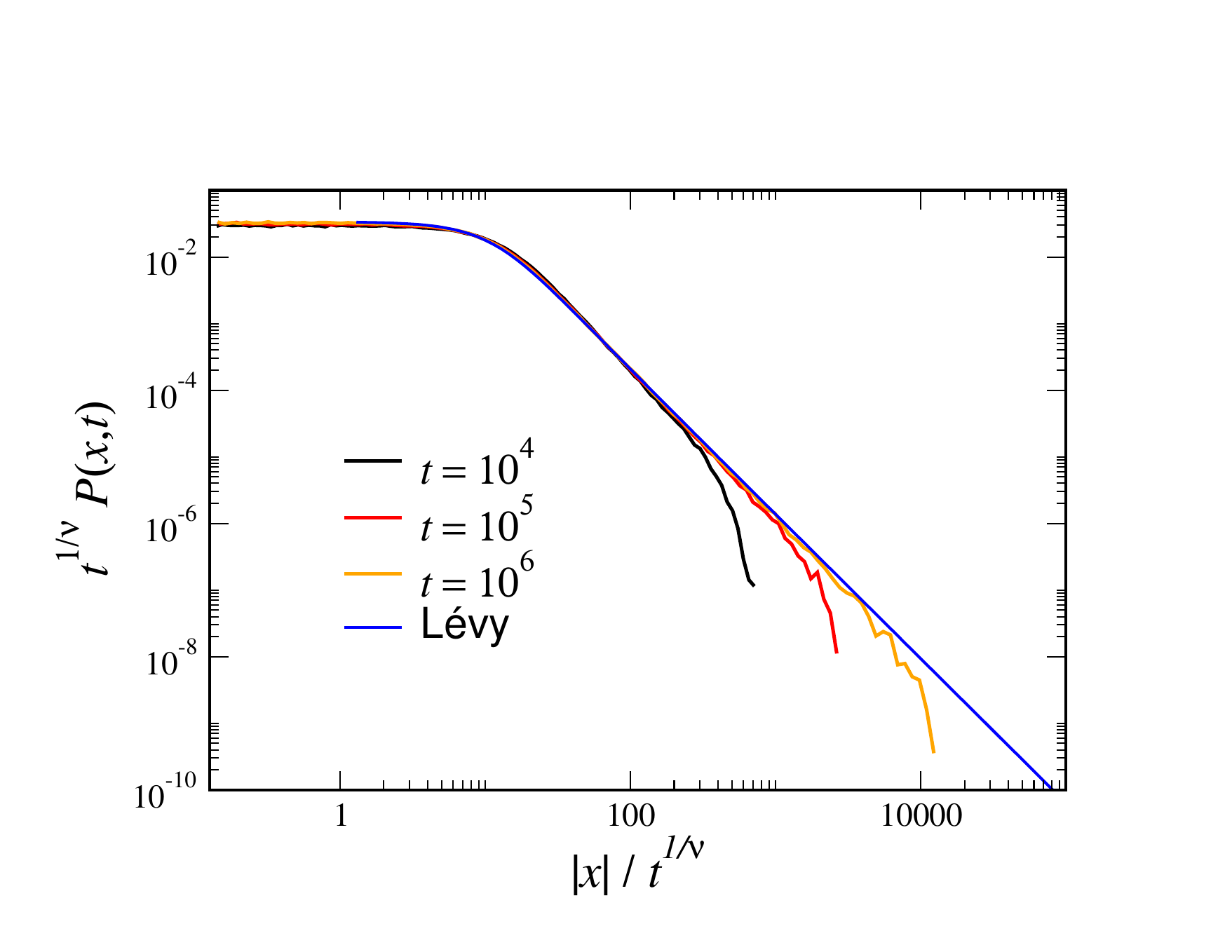,width=2.1in}\label{ra_fig2b}}\\
  \subfigure[]
     {\epsfig{figure=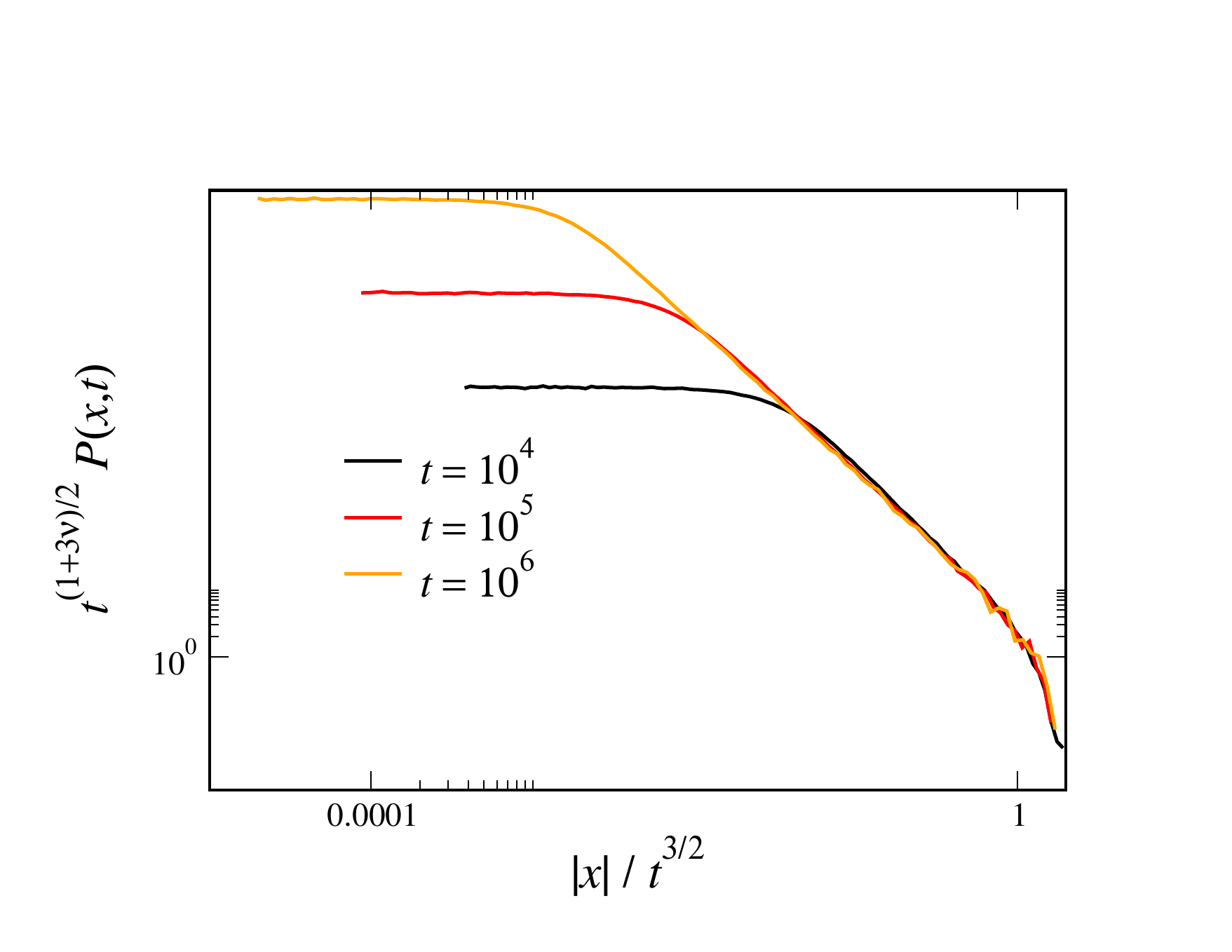,width=2.1in}\label{ra_fig2b}}
\end{center}
\caption{ 
(a)
$t^{1/\nu} P(|x|,t)$ versus $|x|/t^{1/\nu}$ for $D=2/5$, i.e. $\nu=7/6$. The theory, i.e., the
L\'evy PDF Eq. (\ref{eq18}) along with $K_\nu$, Eq. (\ref{eq21}), perfectly matches
simulations without fitting.  (b)  Same data in log-log scale.
Notice the cutoff for large $x$ which is due to the coupling between jump
lengths and jump durations, making $x>t^{3/2}$ extremely improbable.
(c)  $t^{(1+3\nu)/2} P(|x|,t)$ versus $|x|/t^{3/2}$ for $D=2/5$,
 showing the  crossover
from power-law to Gaussian behavior at $|x| \sim t^{3/2}$.
This cutoff ensures the finiteness of the mean square displacement 
of the underlying
spreading of the packet of atoms. 
}
\label{fig3}
\end{figure}

Fig. \ref{fig3} (b) also illustrates the  cutoff on the L\'evy distribution, which is found at distances $x\sim t^{3/2}$, as can be seen also
 from Fig \ref{fig3}(c). 
 Beyond this length scale, the density falls off rapidly (Fig. \ref{fig3}(c)).
  This, as noted above, is the result of the correlation between $\chi$ and $\tau$, as there are essentially no walks with a displacement greater than  of order $t^{3/2}$.  This cutoff ensures the finiteness
of the mean square displacement, using the power law
tail of the L\'evy PDF $L_\nu(x) \sim x^{- (1 + \nu)}$
 and the cutoff   we get:
 $$\langle x^2 \rangle\simeq \int^{t^{3/2}} t^{-(1/\nu)} (x/t^{1/\nu})^{ - (1 + \nu)} x^2 {\rm d} x\sim t^{4 -3 \nu/2},$$
 for $2/3 < \nu < 2$, in agreement with
rigorous results of Ref. \refcite{DLKBprl} (where the prefactor is also computed).
As noted in the introduction, if we rely on the fractional diffusion equation, 
Eq. (\ref{eq01}),
naively, we get $\langle x^2 \rangle=\infty$.
Thus the fractional equation must be used with care,
realizing its limitations in the statistical description of the moments
of the distribution and its tails.

\section{Richardson Phase}
When
 the average jump duration, $\langle \tau \rangle$, diverges,  i.e., for $D>1$,
the dynamics of $P(x,t)$ enters a new phase.  Since the Levy index $\nu$ approaches 2/3 as $D$ approaches 1,
$x$ scales like $t^{3/2}$ in the limit.  Due to the correlations, $x$ cannot grow faster than this, so in this regime,
$P(x,t) \sim t^{-3/2} h(x/t^{3/2})$.
This scaling is that of free diffusion, namely momentum diffuses
like $p \sim t^{1/2}$ and hence the time  integral over the momentum scales
like $x \sim t^{3/2}$. Indeed, in the absence of the logarithmic potential,
namely in the limit $D\gg1$ Eq. (\ref{eq05}) gives
\begin{equation}
P\left(x,t\right) \sim \sqrt{ \frac{3 }{ 4 \pi D t^3}} \exp\left[ - \frac{3 x^2 }{ 4 D t^3} \right].
\label{eq22}
\end{equation}
This limit describes the Obukhov model for a tracer particle path in
turbulence, where
the velocity follows a simple Brownian motion \cite{Obukhov,Friedrich}.
 These scaling properties are
related to Kolmogorov's theory of 1941
(see Eq. (3) in Ref. \refcite{Friedrich}) and
to Richardson diffusion~\cite{Rich,DLKBprl} $\langle x^2 \rangle \sim t^3$
.
Eq. (\ref{eq22}) is valid when the optical potential depth is small
since $D\to \infty$ when $U_0 \to 0$.
This limit should be taken with care,
as the observation time must be made large
before considering the limit of weak potential.
In the opposite scenario, i.e. $U_0 \to 0$ before $t \to \infty$,
we expect ballistic motion, $|x| \sim t$,
since then the optical lattice has not had time to make itself
felt \cite{Sagi}.

\section{Relation between first passage times and diffusivity}

 When the variance of $\chi$ is finite, namely $\nu>2$, we get normal
diffusion so that $P(x,t)$ is a Gaussian.
 In that case the spatio-temporal distribution of jump times
and jump lengths decouples $\psi(\chi,\tau) \simeq q(\chi) g(\tau)$. 
The mean square displacement $\langle x^2 \rangle \sim 2 K_2 t$ and
the diffusion constant is 
\begin{equation}
K_2 = \lim_{\epsilon \to 0} \frac{ \langle \chi^2 \rangle }{ 2 \langle \tau \rangle}.
\label{eqDifCon}
\end{equation} 
This equation relates the first passage time properties of the momentum
process and the spatial diffusivity and 
it has the structure of the famous Einstein relation, relating the
variance of jump sizes and the average time to make a jump, 
with a diffusion constant. However here we are dealing with a special limit,
 namely $\langle \tau \rangle$ is the average time  it takes for a particle with initial
momentum $p_i = \pm \epsilon$  to reach
$p=0$ for the first time. This average first passage time vanishes
in the limit of $\epsilon \to 0$. The variance of $\chi$, 
is $\langle \chi^2\rangle = \int_{-\infty} ^\infty \chi^2 q(\chi) {\rm d} \chi$
 namely  the variance of  the area under the 
random first passage process $p(t')$
till its first return to $p=0$, 
and it too
vanishes in the limit $\epsilon \to 0$. Notice that
here $\chi$ may be either positive or negative 
(see Fig. \ref{fig1})  and its mean is zero (since  the particle may start on
$p_i = \pm \epsilon$) with equal probability since the force
$F(p)$ is zero on $p=0$.
Of-course from symmetry $F(p)=-F(-p)$, so it suffices to consider
 $\chi$ for say positive 
areas only. 

 The calculation of $\langle \tau \rangle$ and $\langle \chi^2 \rangle$
is based on backward Fokker-Planck equations
Eqs. (\ref{eqA01},\ref{eqA03})  (see further details
in Ref. \refcite{KesBarPRX}) and we find
\begin{equation}
K_2 = \frac{1 }{ D \cal Z} \int_{-\infty} ^\infty {\rm d} p e^{ V(p) / D} \left[ 
\int_p ^\infty {\rm d} p' e^{ - V(p') /D} p'\right]^2,
\end{equation}
and ${\cal Z} = \int_{-\infty} ^\infty \exp[ - V(p)/D] {\rm d} p$. 
This equation was derived previously using a different approach 
in Ref. \refcite{Hodapp}.
The coefficient $K_2$ diverges as $D \to 1/5$ from below indicating 
the transition to the superdiffusive phase. 
A sharp increase in the diffusion coefficient of an atomic cloud 
as the laser intensity reaches a critical value was 
detected  experimentally by Hodapp et al. \cite{Hodapp}, 
which could be the fingerprint of the transition from Gaussian
to the L\'evy phase. 
We see that the Green-Kubo formula, Eq. 
(\ref{eqGK}), is not the only way to calculate the diffusion constant.
For Markovian processes, with non-linear friction, using
Eq. (\ref{eqDifCon}) and  calculating
the first passage time properties is an alternative. In the
normal diffusion case, the Green-Kubo formalism is so successful that the existence of
an alternate method is only of academic interest.
 However, as shown throughout this current work,
 calculations of first passage time properties of the
velocity process are very useful for the unraveling  of the anomalous
spatial diffusion properties of the system,  where we cannot use the 
standard Green-Kubo formalism nor the Gaussian central limit theorem.
See further work on a generalized Green-Kubo formalism  and non-stationary  momentum
correlation functions in Ref.
\refcite{DechantPNAS}.

\section{Equation for the distribution of the area swept
by a particle until its first passage}
\label{Sec08}

We briefly outline the derivation of the equation for the PDF of
the  area $\chi = \int_0 ^t p(t') {\rm d} t'$
under the stochastic process $p(t)\ge 0$ which starts on $p_i>0$ 
until its reaches $p=0$ for the first time, so that $t$ is the
first passage time.  The PDF $q(|\chi|)$
has a Laplace $\chi \to s$ transform which we call $Q(s,p_i)$
(here $\chi>0$ the extension to negative $\chi$ being trivial).
The equation for this quantity is 
\begin{equation}
D\frac{d^2 Q}{dp_i^2} + F(p_i) \frac{dQ}{dp_i} - sp_i Q = 0 \ .
\label{eqQPi}
\end{equation}
For simplicity and for conciseness we consider the case $F(p)=0$,
so $p(t)$ is undergoing Brownian motion without bias.
By definition $\chi_{p_i} = \int_0 ^t p(t') {\rm d} t'$ so
\begin{equation}
Q(s,p_i) = \langle \exp[ - s \int_0 ^t p(t') {\rm d} t'] \rangle_{p_i}
\label{eqDDee}
\end{equation}
where $t$ is the first passage time. 
The starting point is denoted
by the subscript. Now we consider the process $p(t)$ as a random walk
on a lattice with probability $1/2$ to jump from $p(t) \to p(t) + \delta p$
or $p(t) \to p(t) - \delta p$, so $\delta p$ is the lattice spacing.
The constant times between these jumps is $\delta t$ and we consider
the limit $\delta p \to 0$ and $\delta t \to 0 $ while keeping
$D = (\delta p)^2/ 2 \delta t$ finite, this being the Brownian limit. 

 A particle starting with $p_i $ at time $t=0$ can either have momentum
$p_i-\delta p$ at time $\delta t$ with probability one half, or  $p_i +  \delta p$
with the same probability. From the definition of $\chi_{p_i}$ and for a particle
starting with $p_i $ we have
\begin{equation}
\chi_{p_i} = \left\{
\begin{array}{l  l } 
p_i   \delta t + \chi_{p_i - \delta p}\qquad & \mbox{Prob} \ \  1/2 \\
p_i  \delta t + \chi_{p_i + \delta p}  & \mbox{Prob} \ \  1/2 .
\end{array}
\right.
\label{eqJumps}
\end{equation}
The first case corresponds to a particle initially jumping down in $p$,
$p_i \rightarrow p_i - \delta p $ while the second case is for a particle
jumping up.  The term $p_i \delta t$
 stems from the displacement
during  the interval $\delta t$, and the remaining term (either $\chi_{p_i}$ or
$\chi_{p_i + 2 \delta p}$  is the displacement for a
new process which starts either with $p_i-\delta_p$ or $p_i +  \delta p$).
 Eq. 
(\ref{eqJumps})
is now used to find
\begin{equation}
 \langle \exp( - s \chi_{p_i}  ) \rangle = 
\frac{1}{2}\sum_{\sigma=\pm 1} 
\langle \exp\left[ - s ( p_i  \delta t - s \chi_{p_i+\sigma \delta p}\right]\rangle.
\label{eqlong}
\end{equation}
Expanding the  right hand side of Eq.  
(\ref{eqlong}) to second order in $\delta p$ and first order in $\delta t$,
and inserting these expansions in Eq. (\ref{eqlong}), using 
$D = (\delta p)^2/ 2 \delta t$,  we get 
Eq. (\ref{eqQPi}). For $F(p_i)\ne 0$ a similar approach is used though now 
weights on the jumping  probabilities must be implemented (see similar
expansions in Refs. \refcite{Kearney,Carmi}). 
The boundary condition $Q(s,p_i)|_{p_i=0}=1$ is used since a particle starting
on $p_i\to 0$ will immediately reach $p=0$ and then the PDF of
$\chi$ is a delta function centered on zero, so the Laplace transform
of this delta function $Q(s,0)$ is equal to unity. 
We use eq. (\ref{eqQPi}) in Appendix A (and in more detail in Ref. \refcite{KesBarPRX})
to derive the long tail of $q(\chi)$ and from it $K_\nu$. 

\section{Discussion of the experiment}
Our work shows a rich phase diagram of the dynamics, with two transition
points. For deep wells, $D<1/5$,
 the diffusion is Gaussian, while for $1/5<D<1$
 we have L\'evy statistics, and for $D>1$
the Richardson-Obukhov scaling  $x \sim t^{3/2}$  is found.
So far,  experiments detected a Gaussian phase and an anomalous
one.  A transition to Richardson-Obukhov scaling
has  not yet been  detected.
Sagi et al. ~\cite{Sagi}  clearly demonstrated that
changing the optical potential depth
modifies the anomalous exponents in the L\'evy spreading packet.
 However, the experiment
showed at most ballistic behavior, with the spreading exponent $\delta$, defined by $x \sim t^\delta$, always less than unity. This might be related to our
observation that to go beyond ballistic motion, $\delta>1$, one
must take the measurement time to be very long.
A more serious problem is that, in the experimental fitting of the diffusion
front
to the L\'evy propagator,
an additional exponent was introduced
 \cite{Sagi}
to describe the time dependence of the full width at half
maximum. In contrast, our semi-classical theory shows that a single
exponent $\nu$ is needed within the L\'evy scaling regime $1/5<D<1$, with the spreading exponent $\delta=1/\nu$.
This might be related to the cutoff of the tails of L\'evy
PDF which demands that the fitting be performed in the central 
part of the atomic cloud.
 On the other hand we cannot
rule out other physical effects not included in the semiclassical model. For example
it would be very interesting to simulate the system with quantum Monte Carlo
simulations, though we note that these are not trivial in the
$|x| \sim t^{3/2}$ regime since the usual simulation procedure
introduces a cutoff on
momentum, which may give rise to an artificial ballistic motion. 
Further, the experiment suffers from leakage of particles, 
which could be crucial, while our theory assumes conservation
 of the number of particles. 
Another effect our theory does not address is the Doppler cooling which may serve as
a cutoff at long time scales (see some discussion on this
in \cite{DechantPNAS}). 
  Thus while there is some tantalizing
points of contact between the theory and experiment, full agreement is yet far away.

\section{Summary}
Usually first passage times are derived from a random diffusive processes.
Here our goal was a bit different. We use first passage time statistics
to derive the spatial diffusivity of the atomic cloud. We revealed
also general relations between first passage times and the normal diffusion
constant $K_2$. Since the latter, through linear response, is in many cases
related to the mobility, we see that analysis of first passage time problems
in momentum space can be very helpful. For this analysis, we need not only distributions
of first passage times (which was a topic
 thoroughly investigated previously) but also
relatively new tools (at least in physics), namely the distributions of areas
under the Langevin curve till the random first arrival, and the analysis of the
area under positive excursion curves like the Bessel excursion with fixed
time interval.
 The area under the Brownian excursion, e.g. the
Airy distribution, is found to be useful in a special
 limit ($D \to \infty$) which 
is encouraging since this mathematical object 
had very little to do with physical behavior up until the recent works
in the field. 
The rich phase diagram, i.e. Normal, L\'evy, Richardson diffusions, together
with
the divergence of the diffusion constant close to the transition, 
revealed by the semi-classical approach used here,
point to interesting physics, though
many open questions  and challenges both experimental
and theoretical remain open with respect to
the dynamics and thermodynamics of laser cooled atoms. 

\section*{Acknowledgements}
This work was supported by the Israel Science Foundation. We thank Andreas
Dechant,
 Eric Lutz, Nir Davidson, Yoav Sagi, and Satja Majumdar for discussions. 
EB thanks the Free University of Berlin for hospitality during the preparation
of the manuscript.

\begin{appendix}[]

In this appendix we derive the small $k$, small $u$ behavior of $\widetilde{\psi}$.
 We first note~\cite{Blumen} that for $u^{3/2} \ll k \ll 1$, $D<1$, we can approximate $\widetilde{\psi}$ by its decoupled form, $\widetilde{\psi}(k,u) \sim \tilde{g}(u) \tilde{q}(k)$.
For small $u$, since $g$ is a probability distribution, we have $\tilde{g}(u)\sim 1 - u\langle \tau \rangle$, as long as the first moment exists.  This first moment
satisfies the backward Fokker-Planck equation~\cite{Gardiner}
\begin{equation}
D\frac{d^2 \langle \tau \rangle}{dp_i^2} + F(p_i) \frac{d \langle \tau \rangle}{dp_i} = -1
\label{eqA01}
\end{equation}
and the small $p_i$ limit of the solution is:
\begin{equation}
\langle \tau \rangle \sim \frac{p_i}{D}  \int_0^\infty e^{-V(p)} dp = \frac{p_i \sqrt{\pi}\Gamma\left(\frac{3\nu - 2}{2}\right)}{2 D\Gamma\left(\frac{3\nu-1}{2}\right)}
\label{eqA02}
\end{equation}
For small $k$ behavior of $\tilde{q}(k)$, given that the second moment diverges, we have $\tilde{q}(k) \sim 1 - \pi x_*^\nu |k|^\nu/ (\Gamma(1+\nu)\sin \frac{\pi \nu}{2})$.  The length scale $x_*$ governing the decay of $q(\chi)$ can in principle be determined from the scaling function $B$, but it is easier to get it directly
from another backward Fokker-Planck equation for the Laplace transform of $q(\chi)$, $Q(s,p_i)$, where we have made explicit the dependence on the initial momentum $p_i >0$ (so $\chi>0$):
\begin{equation}
D\frac{d^2 Q}{dp_i^2} + F(p_i) \frac{dQ}{dp_i} - sp_i Q = 0 \ .
\label{eqA03}
\end{equation}
From this, we derive the small $p_i$, small $s$ behavior:  $Q(s,p_i) \sim 1 - p_i \Gamma(1-\nu)(3\nu-1)^\nu s^\nu /(\Gamma(\nu) 3^{2\nu-1})$, which in turn yields
\begin{equation}
x_*^\nu \sim p_i \frac{\nu(3\nu-1)^\nu}{2\cdot 3^{2\nu-1}\Gamma(\nu) }
\label{eqA04}
\end{equation}
where the $1/2$ prefactor stems from equal consideration of 
positive and negative excursions. 
Putting these two results together, we have
\begin{equation}
\widetilde{\psi}(k,u) \approx 1 - \langle\tau\rangle u - x_*^\nu|k|^\nu\cdots .
\label{psiku}
\end{equation}
To leading order, we may approximate $\Psi(k,u)\sim \Psi(0,u) \sim 1/\langle \tau \rangle$, which, together with Eq. (\ref{psiku}), gives us Eq. (\ref{Levyku}).
\\

\end{appendix}

\bibliographystyle{ws-rv-van}
\bibliography{bibfile}

\end{document}